# Evidence of the inverse proximity effect in tunnel magnetic Josephson Junctions

R. Satariano[1,2*], A. F. Volkov[3], H. G. Ahmad[1], L. Di Palma[1], R. Ferraiuolo[1,2], Z. Iqbal[1], A. Vettoliere[4], C. Granata[4], D. Montemurro[1,2], L. Parlato[1,2], G. P. Pepe[1,2], F. Tafuri[1], G. Ausanio[1,2], and D. Massarotti[5,2]

[1] Dipartimento di Fisica Ettore Pancini, Università degli Studi di Napoli Federico II, c/o Complesso Monte Sant'Angelo, via Cinthia, I-80126 Napoli, Italy.

[2] Consiglio Nazionale delle Ricerche - SPIN, c/o Complesso Monte Sant'Angelo, via Cinthia, I-80126 Napoli, Italy.

[3] Institut für Theoretische Physik III, Ruhr-Universität Bochum, D-44780 Bochum, Germany.
[4] Consiglio Nazionale delle Ricerche—ISASI, Via Campi Flegrei 34, I-80078 Pozzuoli, Italy.
[5] Dipartimento di Ingegneria Elettrica e delle Tecnologie dell'Informazione, Università degli Studi di Napoli Federico II, I-80125 Napoli, Italy.

*Corresponding author e-mail: roberta.satariano@unina.it.

Magnetic Josephson Junctions (MJJs) are a special class of hybrid systems where antagonistic correlations coexist, thus providing a key for advances in weak superconductivity, superconducting spintronics and quantum computation. So far, the memory properties of MJJs have been mostly investigated in view of digital electronics and for spintronic devices at liquid-helium temperature. At the operating temperature of quantum circuits, a magnetic order can rise in a Superconductor (S) at the S/Ferromagnet (F) interface, i.e., the *inverse proximity effect* (IPE), thus leading to a significant modification of the magnetic field patterns in MJJs. In this work, we have carried out a comparative investigation of the magnetic behavior of tunnel MJJs with a strong ferromagnetic layer inserted in the layout of both Nb and Al JJs, respectively. The comparative analysis validates the crucial role of the temperature, the fundamental scaling energies of S/F coupling systems, and the transparency of the S/F interface. This investigation points out that the IPE is a key aspect to consider when designing tunnel MJJs operating well below 4 K and thus in the perspective of hybrid superconducting quantum architectures.

## 1 Introduction

Magnetic Josephson junctions (MJJs) have established a unique playground to explore the interplay between superconductivity and ferromagnetism and to reveal new physical phenomena at Superconductor (S)/ Ferromagnet (F) interface [1-3]. As result of the exchange splitting of the singlet Cooper pairs, the phase difference in MJJs oscillates between 0 and π as a function of the thickness of the ferromagnetic material [4-7]. Spin-triplet pairing can be generated by introducing some non-collinear magnetic homogeneities at the S/F interface, e.g., spin-mixer layers, thus mediating spin polarized dissipationless current [8-11]. Due to their unique physics, MJJs have emerged as a promising platform for superconducting spintronics [12,13], digital electronics [14] and quantum computation applications [15]. So far, their use in superconducting quantum architectures has been mainly proposed for realizing π-phase shifters or *quiet* qubits [16-19]. Recently, the low coherence times of a hybrid ferromagnetic zero-flux-biased flux qubit have been ascribed to dissipation mechanisms occurring in the metallic ferromagnetic barrier [20]. Recent advances in exploiting stacked multilayer with insulating barrier (I) (SIsFS or SIFS JJs [21-25]) or intrinsic insulating ferromagnetic ($I_F$) materials (SI$_F$S JJs) [26,27] have allowed to engineer MJJs with very low damping, even accessing the Macroscopic Quantum Tunnel (MQT) regime at very low temperatures [28]. The integration of tunnel MJJs as active components in quantum circuits is thus gaining more and more attention [29]. The notion of a hybrid ferromagnetic transmon qubit, the so-called *ferro-transmon*, has been proposed in Ref. [30]. In this layout, MJJs' memory properties allow for an alternative control of the qubit frequency through magnetic field pulses with a potential major impact on the scalability of superconducting quantum systems [30].

So far, the functionality of MJJs as magnetic switches for digital electronics [14] and for spintronic devices [13] has been demonstrated mostly at liquid-helium temperature. However, at the operating temperature of quantum circuits, the *inverse proximity effect* (IPE), i.e., the leakage of a ferromagnetic order into a superconductor at the S/F interface, can emerge and lead to a significant modification of the response of the device to an applied field $H$ [31,32]. This phenomenon is a result of the polarization of the Cooper pairs at the S/F interface: the electrons with the spin aligned along the exchange field can easily penetrate into the F layer, while the electrons with the opposite spin tend to stay in S [1,33]. The size of this spin-polarized region is of the order of Cooper pair size, i.e., the superconducting correlation length $\xi_S$ ($\sim 10 - 100$) nm. In the diffusive limit, the magnetic moment $m_{SC}$ induced in S has an opposite direction to the F magnetic moment $m_F$. Under certain conditions, the total magnetic moment in the S region compensates the total magnetic moment of the F film, resulting in a full spin screening [1,34,35].

The full screening of the magnetic moment in F at the S/F interface can affect the magnetic dependence of the critical current $I_c(H)$ in MJJs [36]: the displacement due to the S-magnetization cancels out with that due to the F-magnetization, thus resulting in zero-centered $I_c(H)$ curves [36]. Moreover, also a significant broadening of the central Fraunhofer peak is predicted in SFS JJs [36]. Since the magnetic hysteretic behavior of the critical current is a crucial requirement for the development of a ferro-transmon qubit, we intend to gain a deeper understanding of the memory properties of a SIsFS JJs, characterized by very low damping, down to the temperature $T = 10$ mK. In view to scale circular MJJs to the nanoscale, thus achieving Josephson energy level compatible with typical transmon values [37], strong ferromagnets with significant high remanent magnetization and in-plane anisotropy are required to preserve the two-state magnetic behavior of the critical current. These requirements can be satisfied by using NiFe alloys, which show also desirable switching properties with relatively low coercive fields [38,39].

In this work, we have carried out a comparative investigation of the magnetic field patterns of SIsFS JJs with a strong ferromagnetic $Ni_{80}Fe_{20}$ (Permalloy: Py) barrier based on Nb and Al technology, which are the main superconducting materials for quantum devices [40-43]. We show that in different SIsFS with the same ferromagnet a fundamental role is played by the S/F interface. In the SIsFS JJs where S is Nb, the direct coupling of a strong and thin ferromagnet to the Nb layer allows for the fingerprints of the IPE to emerge in the magnetic field patterns. In contrast, in SIsFS with Al electrodes a poor transparency at the S/F interface enables a complete recovery of the standard behavior of the $I_c(H)$ curves in SIsFS JJs, thus suggesting that these MJJs are robust candidates for hybrid superconducting quantum. This study points out that the inverse proximity effect is a key aspect in designing MJJs in the ferro-transmon perspective.

## 2 Experiment

**Sample fabrication and materials characterization**

We have first characterized the magnetic properties of continuous $Ni_{80}Fe_{20}$ (Permalloy: Py) films with a vibrating sample magnetometer (VSM) by applying an in-plane magnetic field. Details on the set-up can be found in Ref. [31]. Since the underlayer can affect the magnetic properties of a F film in view of elastic strains, surface anisotropy, or magnetostatic interactions [44], to reproduce the S/F interface present in the SIsFS JJs we have deposited thin films of Nb (30 nm)/Py (3 nm) via dc magnetron sputtering on a 1 cm x 1 cm Si/SiO$_x$ substrate. At $T = 12$ K, i.e., above the critical temperature of the Nb $T_c$, the paramagnetic signal of the Nb layer is negligible and we have measured the magnetic signal from the F layer (black curve in Fig. 1a). From Fig. 1a, we can estimate a value of the saturation magnetization $\mu_0 M_s \approx 800$ kA/m comparable with previously published data [45]. To investigate the effect of patterning of the films on magnetic properties, we have realized square Py dots with a lateral size of 5 $\mu$m by optical lithography and lift-off procedure. The elements have been arranged on a square lattice with a period of 15 $\mu$m so that the dipolar interaction between them is negligible [46]. In this way, the measured hysteresis loop of the overall array is the superposition of the magnetization loop of a single element [47]. As a result of the larger demagnetizing [46], the Py dots show an increase of the saturation field $\mu_0 H_s$ up to 200 mT. Nevertheless, the coercive field remains of the same order of magnitude of the corresponding continuous film ($\mu_0 H_c \sim 10$ mT), thus suggesting that for these dimensions the reversal still occurs because of a complex domain wall nucleation-propagation mechanism [47].

In Fig. 1b, we show the hysteresis loop of Py dots in the temperature and magnetic field range of transport measurements on MJJs reported in the next Section. Since the thickness of the s-interlayer is much smaller than the London penetration depth $\lambda_L$ ($\sim 100$ nm [48]) the Nb does not enter the Meissner state and we have just subtracted the paramagnetic signal from the Nb layer to derive the hysteresis loop of the Py. These measurements clearly show that at cryogenic temperatures the hysteresis loop of the Py is temperature-independent.

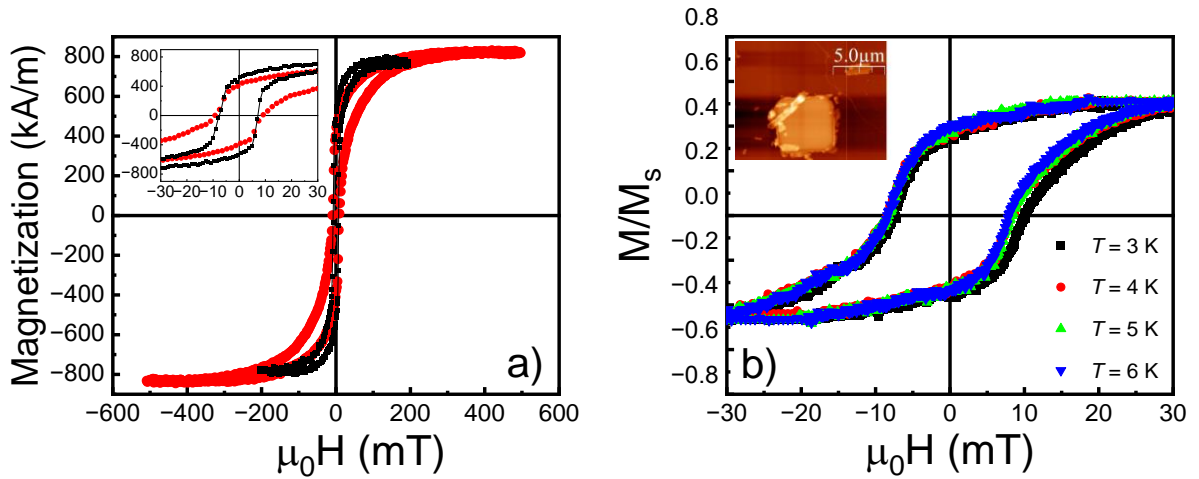

**Fig. 1:** (**a**) Hysteresis loops of a 3 nm-thick continuous Py film (black line) and of a 3 nm-thick Py square dot with lateral size of 5 $\mu$m (red line) at $T = 12$ K. Inset: hysteresis loops at low field. (**b**) Hysteresis loops of a 3 nm-thick Py square dot with lateral size of 5 $\mu$m (red line) measured at different temperatures. Inset: AFM images of some 3 nm-thick Py dot with a lateral size of 5 $\mu$m. In all the panels, the Py has been grown on a 30 nm-thick Nb layer.

To realize SIsFS JJs, we employ a multi-step fabrication process to avoid deteriorating the tunnel properties of the SIs side and contaminating the vacuum chambers of superconducting and ferromagnetic materials. After patterning the SIs trilayer using optical lithography and lift-off procedure, we have defined the junction areas using a selective anodization process followed by a further insulation via SiO$_2$ deposition. The nominal dimension of the diameter of the JJs ranges from 3 to 4 $\mu$m. Then, with vacuum breaking, we have cleaned the surface of s layer with a soft etching by using an ion gun. We have then deposited a 3 nm-thick Py film with a dc magnetron source at a rate of 0.7 nm/s at room temperature. Finally, a S-counter electrode has been deposited by a further dc-sputtering and lift-off process obtaining the overall SIsFS structure. More details on the fabrication process on SIsFS based on Nb and Al technology can be found in Ref. [24] and [49], respectively.

To investigate different transport properties of SIsFS JJs based on Nb technology, we have set the F-thickness $d_F$ at 3 nm and we have decided to prepare two sets of SIsFS JJs with two different s-thickness $d_s$: $d_s = 30$ nm ~ $3\xi_S$ [Nb (200 nm)/Al-AlO$_x$/Nb (10 nm)/Py (3 nm)/Nb (350 nm)] and $d_s = 10$ nm ~ $\xi_S$ [Nb (200 nm)/ Al-AlO$_x$/Nb (30 nm)/Py (3 nm)/Nb (350 nm)]. We have then reproduced the latter layout also for the SIsFS JJs based on Al technology [Al (200 nm)/ Al-AlO$_x$/Al (30 nm)/Py (3 nm)/Al (350 nm)]. As a reference, for each fabrication run of the SIsFS, we have always fabricated a standard SIsS JJs batch from the same wafer by excluding the F layer deposition step.

## Measurements of the magnetic Josephson Junctions

The SIsFS have been measured by thermally anchoring the samples to the mixing chamber of a Triton dry dilution refrigerator provided by Oxford instruments, with a base temperature of about 10 mK. Details on various stages of filtering and on low-noise electronics are reported elsewhere [18]. A superconducting NbTi coil is used to apply uniform in-plane magnetic fields.

In Fig. 2a the magnetic field pattern at $T = 10$ mK of a standard circular SIsS Nb (200 nm)/ AlO$_x$/ Nb (10 nm)/ Nb (350 nm) JJ with radius $R = 1.5$ μm is reported. The $I_c(H)$ curve has been fitted by considering the Airy pattern as the functional form [50]:

$$I_c = I_{c,max} \left| \frac{2j_1\left(\frac{2\pi\Phi}{\Phi_0}\right)}{\frac{2\pi\Phi}{\Phi_0}} \right|, (1)$$

where $j_1$ is a Bessel function of the first kind and $\Phi_0 = \frac{h}{2e}$ is the magnetic quantum flux. The magnetic flux through the junction is $\Phi = \mu_0 H 2R (d_I + 2\lambda_L)$, where $d_I$ is the thicknesses of the I layer and $\lambda_L$ is the London penetration depth. The fitted values $R = 1.52 \pm 0.02$ μm and $\lambda_L = 120 \pm 20$ nm are in agreement with nominal junction dimensions and expected magnetic penetration depth for Nb [48] [51]. The excellent fit indicates that the current density is homogeneously distributed throughout the junction [52].

In Fig. 2b, we show the corresponding experimental data for a circular SIsFS Nb (200 nm)/ Al-AlO$_x$/Nb (30 nm)/Py (3 nm)/Nb (350 nm) JJ with $R = 1.5$ μm. Since the s interlayer is sufficiently thicker than the superconducting coherence length $\xi_S$ (~10 nm [53]), the current-voltage (I-V) characteristic, reported in the inset in Fig. 2b, is typical of a tunnel junction with $I_c R_N = 1.2$ mV, where $R_N$ is the normal state resistance, which is reduced by only 20% with respect to the reference SIsS JJ ($I_c R_N = 1.5$ mV from the I-V curve in the inset in Fig. 2a). Therefore, the device can be considered as a SIs junction in series with a sFS junction: the SIs JJ with the smaller critical current sets the transport behavior of the overall structure [54], since the phase difference mainly drops across the tunnel barrier, while the phase shift at the sFS side is negligibly small.

On the other hand, since $d_s < \lambda_L$, the s-thickness is sufficiently thin to not screen magnetic field and the whole structure behaves as a single junction with respect to an external magnetic field $H$. If $M_F$ is uniform and is oriented as in the same direction of the applied field $H$, the total flux trough the junction becomes:

$$\Phi = \mu_0 H 2R d_m + \mu_0 M_F 2R d_F, (2)$$

where the thickness of the material penetrated by the applied field is $d_m = 2\lambda_L + d_s + d_F + d_I$, with $d_F$ the thickness of the intermediate F-layer [22].

The Airy pattern will be shifted in field by an amount:

$$\pm\mu_0 H_{shift} = \mp \frac{\mu_0 M_F d_F}{d_m}. (3)$$

As a result, along the falling sweeping field direction (*down curve*), the Fraunhofer-like pattern is expected to be shifted at a negative field, because of the positive remanence of the ferromagnet. Along the rising sweeping field direction (*up curve*), the opposite holds.
However, in Fig. 2b and also for different sweeping field range below 40 mT, not shown here, we can observe two main anomalies: i) there is a lack of hysteresis of the $I_c(H)$ curves and ii) the central peak in the SIsFS JJs is two times as wide as the corresponding non-magnetic SIsS JJs.

Regarding the latter point, for a standard Airy pattern, the minimum is expected at $\mu_0 H_{min} = (1.22\Phi_0)/d_m$ [50]. Therefore, for a MJJ, the half-width of the central peak $\mu_0 \Delta H = \mu_0(H_{min} - H_{shift})$ should not depend on additional flux, but only on geometrical parameters of the JJs. Moreover, the F-layer's domain structure usually manifests itself in narrower central peaks and in shifts of the $I_c(H)$ curves approaching the coercive field [56].

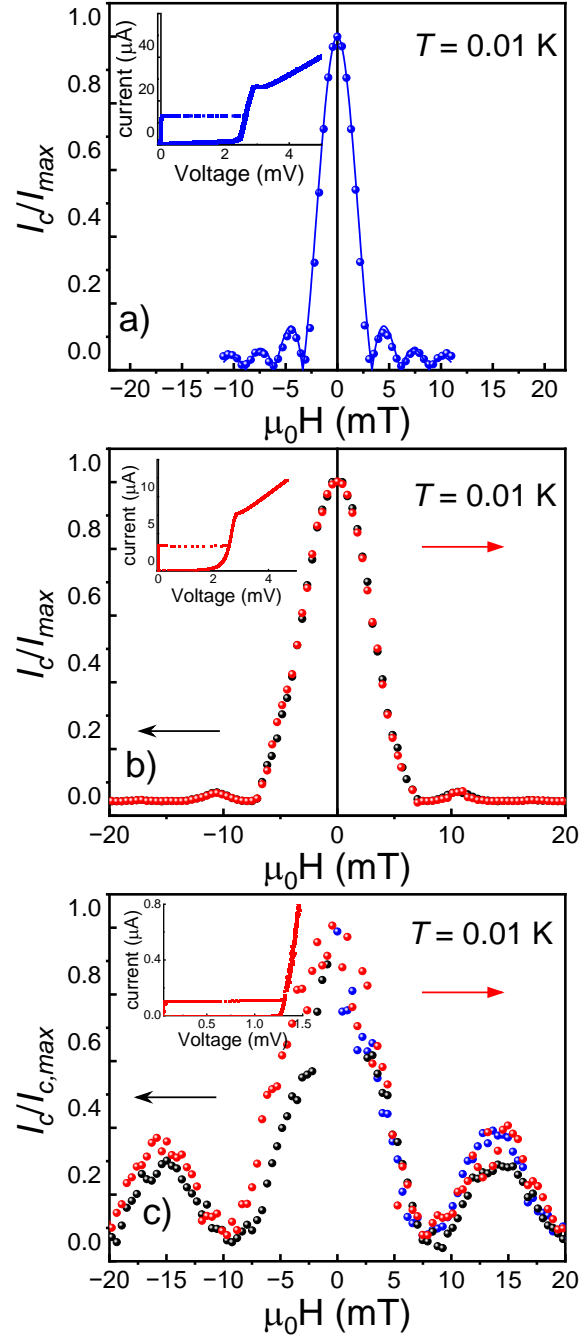

**Fig. 2.** Normalized $I_c$ as a function of the magnetic field $H$ for (**a**) a circular SIsS [Nb (200 nm)/AlO$_x$/Nb (10 nm)/Nb (350 nm)] JJ with radius $R = 1.5$ μm, (**b**) a circular SIsFS [Nb (200 nm)/ Al-AlO$_x$/Nb (30 nm)/Py (3 nm)/Nb (350 nm)] JJ with $R = 1.5$ μm, and (**c**) a circular SIsFS [Nb (200 nm)/ Al-AlO$_x$/Nb (10 nm)/Py (3 nm)/Nb (350 nm)] JJ with $R = 2$ μm. In both panels b and c, the black and red curves are the magnetic patterns in the downward and upward direction of the magnetic field, respectively. In panel c, the blue curve is the magnetic field pattern for the virgin state of the ferromagnet. The arrows indicate the sweeping field directions. For each panel, the current-voltage characteristic at zero field has been shown in the inset. All the magnetic field patterns are measured at temperature $T = 10$ mK and have been normalized to the maximum value of the critical current $I_{max}$.

As shown in the following, the lack of hysteresis of the $I_c(H)$ curves cannot be naively ascribed to the fact that the SIsFS in Fig. 2b behaves as a serial connection of a SIs and sFS JJs. In Fig. 2c, we show the magnetic field patterns and I-V curve of a circular Nb (200 nm)/ Al-AlO$_x$/Nb (10 nm)/Py (3 nm)/Nb (350 nm) JJ with $R = 2$ μm. In this case, $d_S \sim \xi_S$ and the superconductivity in the s interlayer is suppressed due to its small thickness and the proximity with a F film. Since the phase drop occurs across the IsF barrier, the critical current density is two orders of magnitude lower as compared to the SIsS JJs (compare the inset in Fig. 2a e 2c). Nevertheless, this junction shows a standard tunnel I-V characteristic at base temperature with a value of the critical current approaching that required for a ferro-transmon. However, even in this different transport regime, we observe a lack of hysteresis of the magnetic field pattern that is not advantageous for the ferro-transmon perspective [30]. Moreover, the central peak is four times larger than the corresponding SIsS JJ.

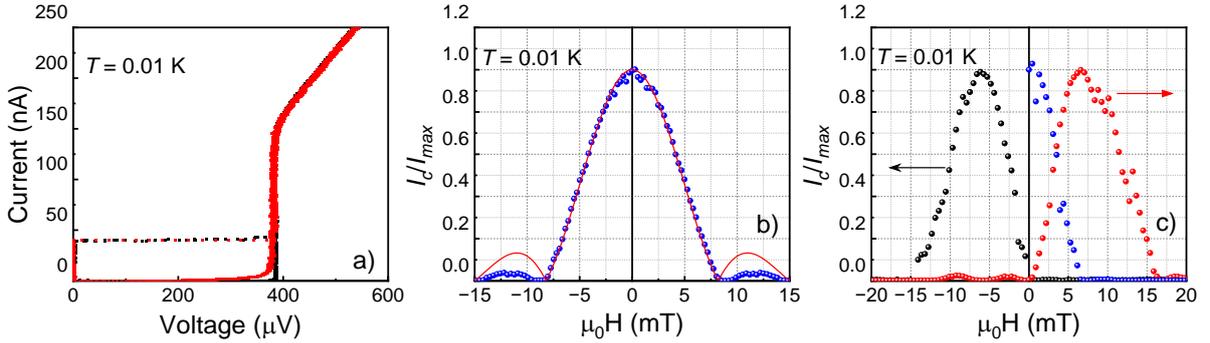

**Fig. 3. a)** Current-voltage curve for a SIsS [Al (200 nm)/ Al-AlO$_x$/Al (30 nm)/Al (350 nm): black curve] and SIsFS [Al (200 nm)/ Al-AlO$_x$/Al (30 nm)/Py (3 nm)/Al (350 nm): red curve)] JJ with a radius $R = 2$ μm. Critical current vs in-plane magnetic field ($I_c(H)$ curves) measured at $T = 10$ mK for **(b)** a SIsS JJ with radius $R = 2$ μm, showing the familiar Airy diffraction pattern (red line) and **(c)** for a SIsFS JJ with radius $R = 2$ μm. The blue curve is the magnetic field pattern for the virgin state of the ferromagnet. The black and the red curves are the magnetic pattern in the downward and upward direction of the magnetic field, respectively. All the magnetic field patterns are measured at temperature $T = 10$ mK and have been normalized to the maximum value of the critical current $I_{max}$.

SIsFS JJs based on Al technology [Al (200 nm)/ Al-AlO$_x$/Al (30 nm)/Py (3 nm)/Al (350 nm)] show a standard behavior of the magnetic field pattern (Fig. 3). For thicknesses greater than ∼ 3 nm, the phenomenological dependence of the Al gap $\Delta$ on thickness $d$ is of the form ($\Delta(d) = \Delta_{bulk} + ad^{-1}$), where the ($\Delta_{bulk} \simeq 180\ \mu V$) is the gap in the bulk limit and $a$ is a parameter that depends on the details of the deposition process [57]. For the set of measurements in Fig. 3a, the pair potential in the s interlayer is even larger than the one in electrodes and the SIsFS can be considered as a serial connection of a tunnel SIs JJ and a ferromagnetic sFS JJ. Since $I_c^{SIs} \ll I_c^{sFS}$, the I-V curve of the overall SIsFS device is determined by its SIs part [54,55]. As a result, the I-V curve of the SIsFS in Fig. 3a basically coincides with the standard non-magnetic one [25,49,58]. The dependence of $I_c$ as a function of $H$ at $T = 10$ mK is shown in Fig. 3b for a circular SIsS JJ with radius $R = 2$ μm, from which the Al London penetration depth can be determined ($\lambda_L \sim 35$ nm $> d_s$). Therefore, a conventional hysteretic behavior of the $I_c(H)$ curves of the magnetic field pattern of the SIsFS JJs is observed in Fig. 3c. In this case, we also observe a semi-width $\mu_0 \Delta H \sim 8$ mT as in Fig. 3b.

To gain some insights into the origin of the unconventional magnetic behavior of the SIsFS based on Nb technology, we have decided to perform the $I_c(H)$ measurement by varying the temperature. The measurements for a circular SIsFS JJ with $d_S = 30$ nm by sweeping the field in the downward direction are shown in Fig. 4a, while the ones acquired in the upward direction are reported in Fig. 4b. For every curve, the values of the shift $\mu_0 H_{shift}$ (Fig. 4c) and of the semi-width $\mu_0 \Delta H = \mu_0(H_{min} - H_{shift})$ (Fig. 4d) have been evaluated. It is clear from Fig. 4c that zero-shifted curves are observed below $T = 4$ K, while above $T = 4$ K the ordinary hysteresis is recovered. As the temperature increases, the shift in field increases.

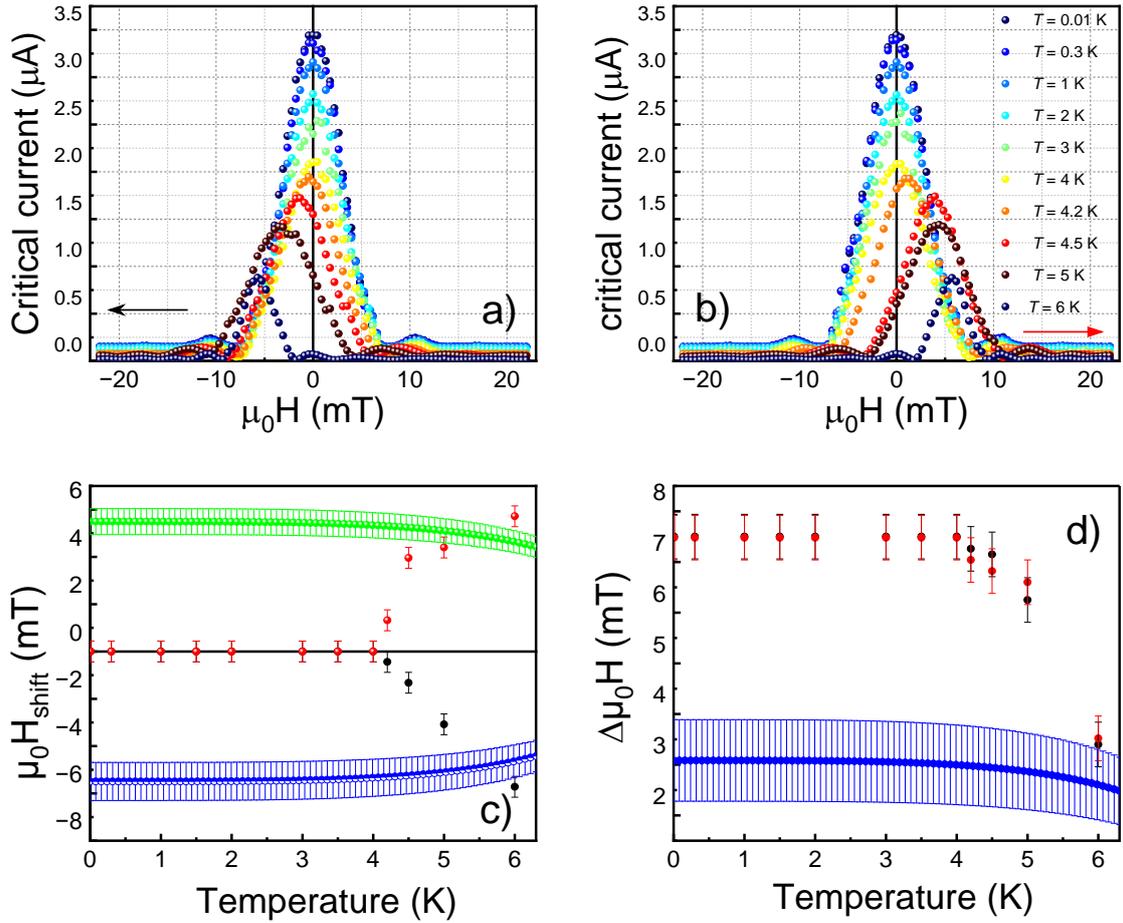

**Fig. 4.** Magnetic dependence of the critical current $I_c(H)$ by sweeping the field in the range (-22, 22) mT is reported in (**a**) in the downward and in (**b**) upward direction of the magnetic field, respectively. The color in the legend indicates the temperature $T$ at which the measurements have been acquired. Temperature dependence of the experimental (**c**) shift and (**d**) semi-width of the magnetic field patterns shown in panel a and b. The black and red dots indicate the experimental data for the downward and upward magnetic field curves, respectively. The blue and green dots indicate the temperature behavior of the shift and semiwidth expected for the down and up curves, respectively, in absence of the inverse proximity effect, as discussed in the text.

## 3 Discussion

The thermal behavior of the magnetic field patterns in Fig. 4 cannot be related to the temperature dependence of the $M(H)$ loops of F-layer, which are temperature independent below 6 K (Fig. 1b). In Fig. 4c, we have derived the temperature dependence of the $\mu_0 H_{shift}$ shift by considering the Equation 3 and the maximum magnetization of the Py dot in Fig. 1b as the $M_F$ value. In Fig. 4d, we plot the semi-width by considering $\mu_0 \Delta H = \frac{1.22\ \Phi_0}{2R d_m}$. For both panel c and d, we consider a Ginzburg-Landau temperature dependence of the London penetration depth $\lambda(T)$ with a value of $\lambda_L = 120 \pm 20$ nm at base temperature, as a result of the fit of the Airy pattern in Fig. 2a (blue and green dots are related to the down and up curves, respectively). The error bars have been calculated by propagating the errors on the magnetic thickness $d_m$ and on the radius $R$ (~10%) due to lithography process. For both parameters, a 20% variation is expected between base temperature and $T$ = 6 K if we only consider the temperature dependence of the London penetration depth $\lambda(T)$. Nevertheless, these features become consistent with the expected values, within the experimental errors, at $T$ = 6 K.

At low temperatures, a full screening regime can emerge in the superconductor due to the inverse proximity effect at the S/F interface. So far, this condition has been overlooked and not fulfilled even in experiments involving local magnetic probe, thus leading to weaker measured signal and elusive results [59-61]. In the full screening regime, in a MJJ the flux due to the s-magnetic moment $m_{SC}$ cancels out the flux due to the F-magnetic moment $m_F$, resulting in a lack of hysteresis of the magnetic field patterns [36]. Within the diffusive limit, the induced magnetic moment $m_{SC}$ depends on the exchange energy $J$, the superconducting gap $\Delta$ and a parameter that accounts for the transport properties of the F layer and the transparency of the S/F interface: $\varepsilon_{b,F} = \hbar D_F / (R_b \sigma_F d_F)$, where $D_F$ is the diffusion coefficient of the F-layer, $R_b$ is the S/F interface resistance per unit area and $\sigma_F$ is the F-conductivity. In Ref. [35], it has been shown that in the limit $\varepsilon_{b,F} > \Delta$ and at very low temperatures, a full screening occurs provided by $J < \varepsilon_{b,F}$. For Nb/Py proximity-coupled system, for which $J / \Delta \sim 10$ [62], the full spin screening at low temperatures is in agreement with the theory if we assume a value of $\varepsilon_{b,F}$ at least equal to $10\Delta$. To estimate this scaling energy, we have measured the resistivity of a 3 nm-thick Py layer by using a four-probe-in-line technique and we have found $\rho_F = 84$ Ωcm. Thus, the diffusion coefficient $D_F$ can be obtained via the relation $D_F = \frac{v_F l_F}{3}$, where $v_F = 2.2 \times 10^5$ ms$^{-1}$ is the Fermi velocity and $l_F$ is the mean free path given by $\rho_F l_F = 31.5 \times 10^{-6}$ $\mu\Omega$ cm$^2$ [63]. For these values, $\varepsilon_{b,F} / \Delta$ of order of 10 corresponds to a value of $R_b$ of the same order of magnitude (fΩm$^2$) of MJJs with Nb/Py interface [64].

As additional fingerprints of the IPE, we observe a broadening of the central peak at low temperatures. In Ref. [36], it has been demonstrated that in SFS JJs the width of the central peak can be directly related to the induced magnetization $m_{SC}$, which is a function of the phase difference $\Psi$ across the SFS. By contrast, we observe a phase drop across the SIs side of the junction: for this reason, calculations on the shape of the magnetic field pattern reported in Ref. [36] cannot be applied to our SIsFS JJs in Fig. 2b. Nevertheless, the evident broadening of the central peak at low temperatures can be qualitatively ascribed in terms of the IPE. Moreover, the pattern's widening is even larger in Fig. 2c, where we are approaching a transport regime in which the phase drop is across the F layer, thus supporting our hypothesis. Based on these experimental findings, we can conclude that the inverse proximity effect and the polarization of the Cooper pairs at the S/F interface is the origin of the unconventional behavior at base temperature.

In contrast, in SIsFS JJs based on Al technology we recover a conventional magnetic hysteresis of the magnetic field patterns, due to the reversal of the Py layer, and a width consistent with the standard SIsS JJs. In these SIsFS JJs, we have used the same F layer and thickness that in Nb based batches. However, for the former case after the definition of the junction, we break the vacuum and a thin oxide layer may form at the S/F interface, thus decoupling the s and F layers [25,49]. This occurrence can lead to a higher $R_b$ and thus a smaller $\varepsilon_{b,F}$, thus resulting in the condition $J > \varepsilon_{b,F}$.

### 4 Conclusion

In this work, we have studied the magnetic field behavior of tunnel MJJs based on Nb and Al electrodes with strong ferromagnetic barriers. The comparative investigation identifies the scaling energies to observe a full screening of the F-magnetic moment at very low temperatures where the physics of MJJs is not commonly explored, along with the role played the S/F transparency. The pursuit of acquiring a complete understanding and control of the leakage of the magnetic order at S/F interfaces is essential in the context of hybrid superconducting quantum devices, where preserving magnetic hysteresis is essential to offer alternative control of the qubit frequency. Careful attention must be paid to prevent the full screening regime by engineering the S/F interface, e.g., by adding some insulating thin oxide or metallic buffer layer.

**Acknowledgments**

This work has been supported by the Pathfinder EIC 2023 project "FERROMON-Ferrotransmons and Ferro-gatemons for Scalable Superconducting Quantum Computers", the PNRR MUR project PE0000023-NQSTI, the PNRR MUR project CN-00000013-ICSC and Programma STAR PLUS 2020, Finanziamento della Ricerca di Ateneo, University of Napoli Federico II. H.G.A., D.Ma. (Davide Massarotti), D.Mo. (Domenico Montemurro) and F.T. thank SUPERQUMAP project (COST Action CA21144).